\documentclass{elsart}
\usepackage{graphics}
\usepackage{graphicx}
\usepackage{epsfig}
\usepackage[latin1]{inputenc}
\usepackage{natbib}
\usepackage{citesort}
\usepackage{amssymb}

\begin{document}

\begin{frontmatter}


%\title{Exponential Recovery of Low Frequency Fluctuations in a Diode Laser with Optical Feedback} %\thanksref{label1}}
%\author{Jhon F. Martinez Avila, Hugo L. D. de Souza Cavalcante and José R. Rios Leite \corauthref{cor1}\thanksref{label2}}
%\ead{jhofrema@df.ufpe.br}
%\corauth[cor1]{Professor}
%\address{Departamento de F\'{\i}sica,~Universidade Federal de Pernambuco \\
%50670-901 Recife, PE, Brazil\thanksref{label3}}

\title{Exponential Recovery of Low Frequency Fluctuations in a Diode Laser with Optical Feedback}


%\author{Jhon F. Martinez Avila, Hugo L. D. de S. Cavalcante  \and J. R. Rios Leite}
\author{Jhon F. Martinez Avila, }
\author{Hugo L. D. de S. Cavalcante \and }
\author{J. R. Rios Leite}

\address{Departamento de F\'{\i}sica,~Universidade Federal de Pernambuco \\
50670-901 Recife, PE, Brazil}

\begin{abstract}
We show that the recovery after each power drop on the chaotic Low Frequency Fluctuations in a semiconductor laser with optical feedback follows an exponential envelope. The time constant for such exponential behavior was experimentally measured. This recovery time constant and the average time interval between consecutive drops are shown to have different dependences when measured as function of the pump current.
\end{abstract}

\end{frontmatter}

% main text
\section{Introduction}
\label{intro}

% \appendix

\begin{figure}[!hbtp]
\begin{center}
\resizebox{7.5cm}{!}{\includegraphics{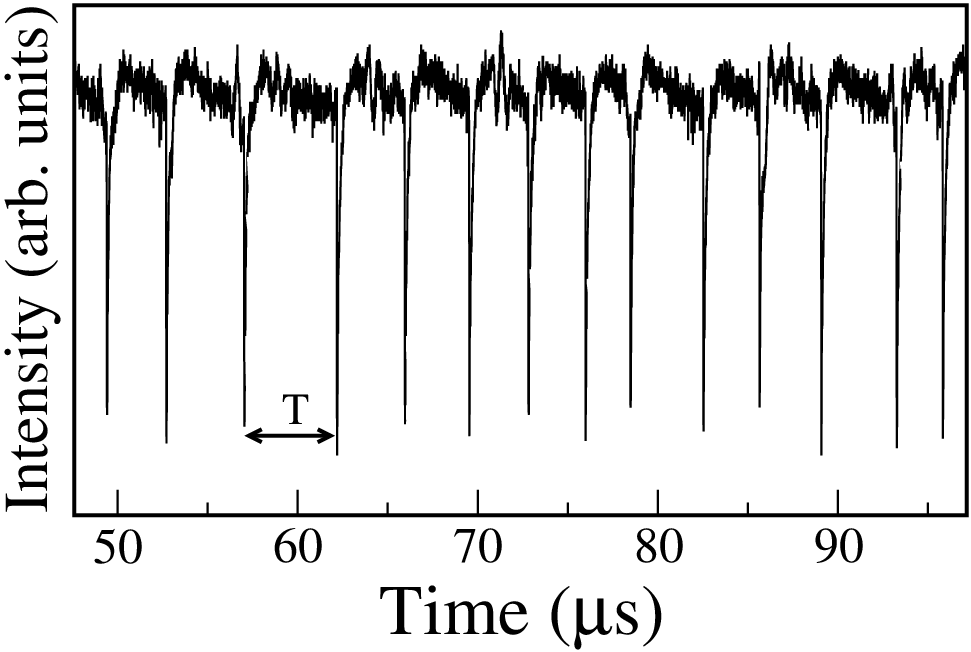}}
\caption{Experimental laser power series with low frequency fluctuation drop
events. The data correspond
to filtered data series with pump current $19\,$mA, 
$\tau$=$30\,$ns and $\xi$=12\,$\%$.}
\end{center}
\label{fig:serielong19ma-30ns}
\end{figure}

% \section{}
% \label{}
Nearly three decades ago, \citet{mork} reported that diode lasers 
submitted to moderate optical feedback show chaotic Low Frequency 
Fluctuations as drops in their power output. 
% Diode lasers,  
%when subject to moderate optical feedback, show chaotic Low 
%Frequency Fluctuations (LFF) as drops in their power
%output as reported by \citet{mork}, nearly three decades ago.
Controlled studies are done using a reflecting mirror and 
making the so called external cavity that feeds back a delayed 
optical field into the very small laser chip cavity.
Thus, the feedback delay time, $\tau$, corresponds to the 
roundtrip time of the external cavity.
The time, $\tau$, is in the range of tens of nanoseconds 
when the reflecting surface is located a few meters from the laser. 
The irregular variation of time interval 
between power  drops, $T$,  
is the main indicator of the chaotic LFF pulsations of the laser, as
seen in figure 
\ref{fig:serielong19ma-30ns}.

%\begin{center} 
%\end{center}

Typically, the average value of  $T$  
changes with the values of the feedback level and the laser pump current, 
going from the order of microseconds down to hundreds of nanoseconds. 
Simultaneously, underneath these slow time scale events, the laser dynamics has very fast 
(picosecond range) pulsations in its  power output and population inversion   
\cite{sano,tartwijklenstra,fischer,vaschenko,sukow99,gavrielides}.
Averaged in a time scale of nanoseconds, the LFF events reported to 
date have the shape of a sharp drop followed by a stepwise
recovery \cite{mork,liudavis,hegartyfast}. 
The deterministic and random contributions to the origin of the LFF 
drops is still subject of studies and many works have been dedicated to the measurement and 
calculation of the properties of $T$ \cite{Hohlroy,wing-shunroy03,sukow97,morkoc99}.
\citet{sukow97} have clearly established that the drops and recoveries within a LFF 
series are nearly constant, while T fluctuates.
\citet{liudavis} studied the recovery process and established that the time interval 
between steps during the LFF recovery correspond to the external round trip or
feedback time delay,  $\tau$. Their work also shows that within LFF time series,
while the time between drops had a wide variance, the number of steps in the recovery 
was nearly constant. They study the number of steps within 
consecutive drops (that is, within the time $T$) as function 
of the current and external cavity length. 
\citet{hegartyfast} also  reported that the fast population relaxation oscillations in diode 
lasers with optical feedback have the same repeated shape while the
interdrop time, $T$, have large chaotic fluctuations. 
They define as recovery time the time interval between 
switch-off and switch-on of the first peak on the ultrafast 
(subnanoseconds) dynamics and show the analogy with the turning-on of the laser. 
However,  
details or measurements for the early stage
after each drop within nanoseconds averaged time are missing in the literature. 
In this time scale the ultrafast highly irregular 
pulses are not detectable and their effect merges 
within the noise that eventually drives the LFF. 
The resulting averaged LFF signal still have important  
physical behavior to be studied quantitativelly as function of its parameters.

The aim of this communication is to define and present quantitative measurements on this 
averaged stepwise recovery of LFF
in a semiconductor lasers with delayed feedback. 
A typical segment of the laser power time series of our experiments, 
filtered with $1$ ns time window, is shown in  figure~\ref{fig:serielong19ma-30ns}.
With the averaged signal we show that the LFF recovery occurs 
with an exponential time dependent envelope, having a time constant $\tau_0$. 
 A recovery time of the LFF events is defined as the time constant of these exponentials. 
This $\tau_0$ is directly measured and is shown to be dynamically 
independent of the time $T$ between drops. 
Such results can be foreseen in figure~\ref{fig:30ldserie-fit}.
To our knowledge, this exponential envelope and its time constant
are proposed and systematically measured for the first time herein.

\section{Experiments}

\begin{figure}[!hbtp]
\begin{center}
\resizebox{7.5cm}{!}{\includegraphics{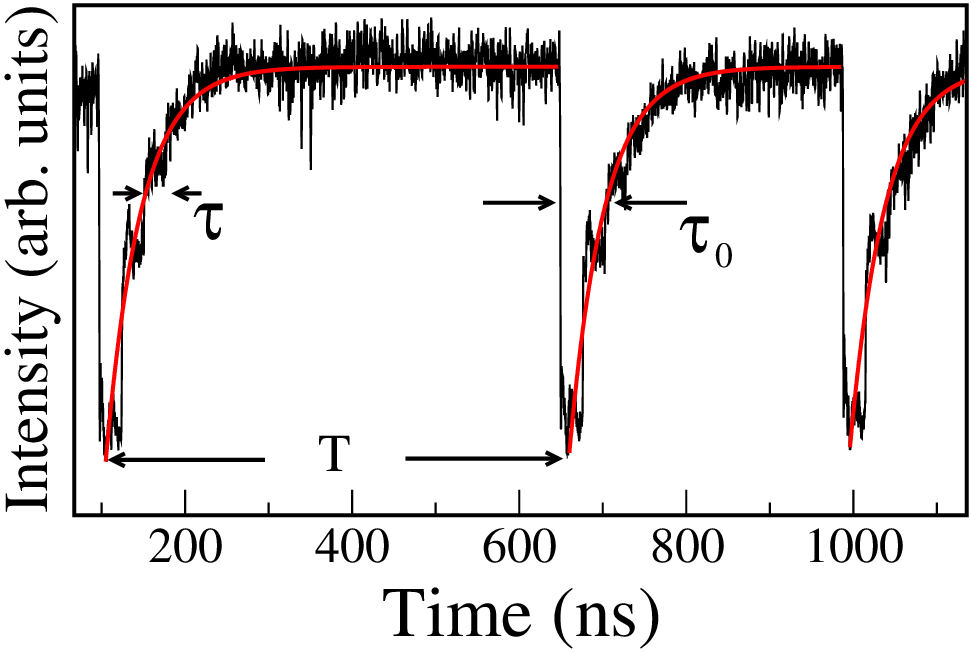}}
\caption{Segment of experimental laser power with three low frequency fluctuation drop
events. The feedback time $\tau$, the interdrop time $T$ and the time 
constant of the recoveries, $\tau_0$, are shown. The smooth (color online) 
curves are exponential fittings using 
Equation \ref{eq:recovertime}. The data correspond
to filtered time series with pump current of $24\, $mA,
$\tau$=$30\,$ns and $\xi$=12\,$\%$. }
\end{center}
\label{fig:30ldserie-fit}
\end{figure}

The laser in the experiments was an SDL 5401  GaAlAs, thermally stabilized
to 0.01 K, and emitting at $850\,$nm. Its solitary threshold current was $17\,$mA.
An external flat mirror distant between $0.9\,$m to $9.0\,$m 
was placed to create the optical feedback. Collimating lenses fixed the
amount of feedback, which was measured by the threshold 
reduction parameter, $\xi$ \cite{tartwijklenstra}.
The intensity output is detected by a 1.5 GHz bandwidth photodiode and the power
data series were filtered with the time averaging of $1\,$ns by a digital oscilloscope.
Very long data series were also stored in a computer memory with
a $12$ bits $A\/D$ acquisition system running at $100$ MHz.
Despite the filtering time of $1\,$ns the use of large round trip time delay
permitted the distinct manifestation of the steps on the recovery of the LFF. 
This is seen in the short segments of the power series of 
figure~\ref{fig:30ldserie-fit} where $\tau$ is $30\,$ns.

Within figure~\ref{fig:30ldserie-fit}
are indicated the external cavity round trip time, $\tau$, 
the time between drops, $T$,
and the recovery time constant of the drop envelope, $\tau_0$. 
To verify phenomenologically the exponential behavior on the 
recovery, after time $t_i$ when a sharp drop occurs, the 
experimental laser power is fitted to the expression
\begin{equation}
P(t)= (P_{0}-P_{L})(1-\exp\{-(t-t_i)/\tau_0\}) + P_{L} ,
\label{eq:recovertime}
\end{equation}
where $P_{L}$ is the minimum value of the laser power, just after 
the drop at time $t_i$ and  $P_{0}$ the value 
just before the drop $(i+1)$.
Figure \ref{fig:30ldserie-fit} shows the fittings of exponentials 
curves (color on line), following equation \ref{eq:recovertime} 
drawn over the experimental data. It is relevant to emphasize that  a
single value of $\tau_0$ was used to fit the three consecutive drops 
presented in figure \ref{fig:30ldserie-fit}. Such is going to be our 
definition of recovery time in a LFF drop event. 

\begin{figure}[!hbtp]
\begin{center}
\resizebox{7.5cm}{!}{\includegraphics{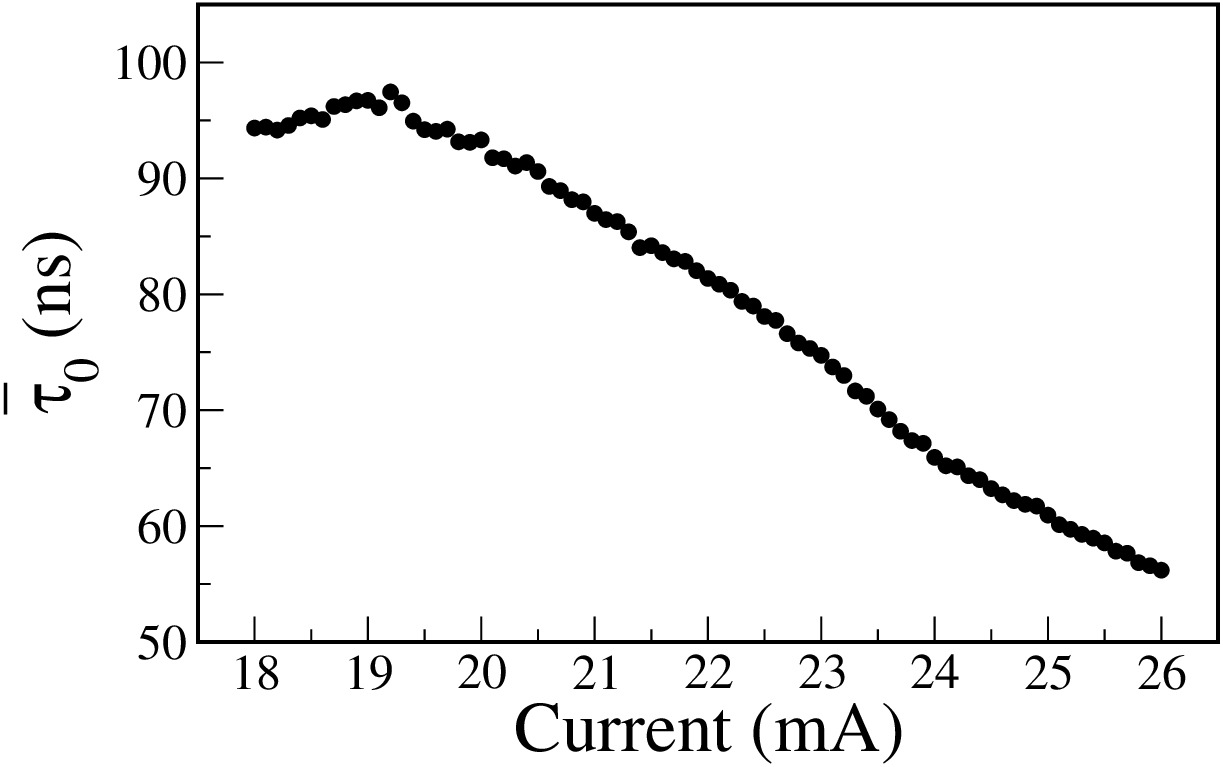}}
\caption{Experimental average of the time constant $\tau_0$ in Low Frequency 
Fluctuating power series, as function of the pump current.
The feedback parameter was $\xi$=12\,$\%$ and the delay $\tau$=$30\,$ns.}
\end{center}
\label{fig:tau0medio30ns}
\end{figure}

For a quantitative study of the experimental value of $\tau_0$, a best
fitting computer program for the parameters of equation 
\ref{eq:recovertime} was  run over long experimental series. 
The program first found the value $t_i$ of a minimum then determined 
$P_{0}$, $P_{L}$ and finally searched for the best $\tau_{0}$. 
From data series having more than $10^4$ LFF drops,
at each value of the injection current,
the computer routine found the average of the recovery time, $\overline \tau_0$, 
which is shown in figure \ref{fig:tau0medio30ns}.

\begin{figure}[!hbtp]
\begin{center}
\resizebox{7.5cm}{!}{\includegraphics{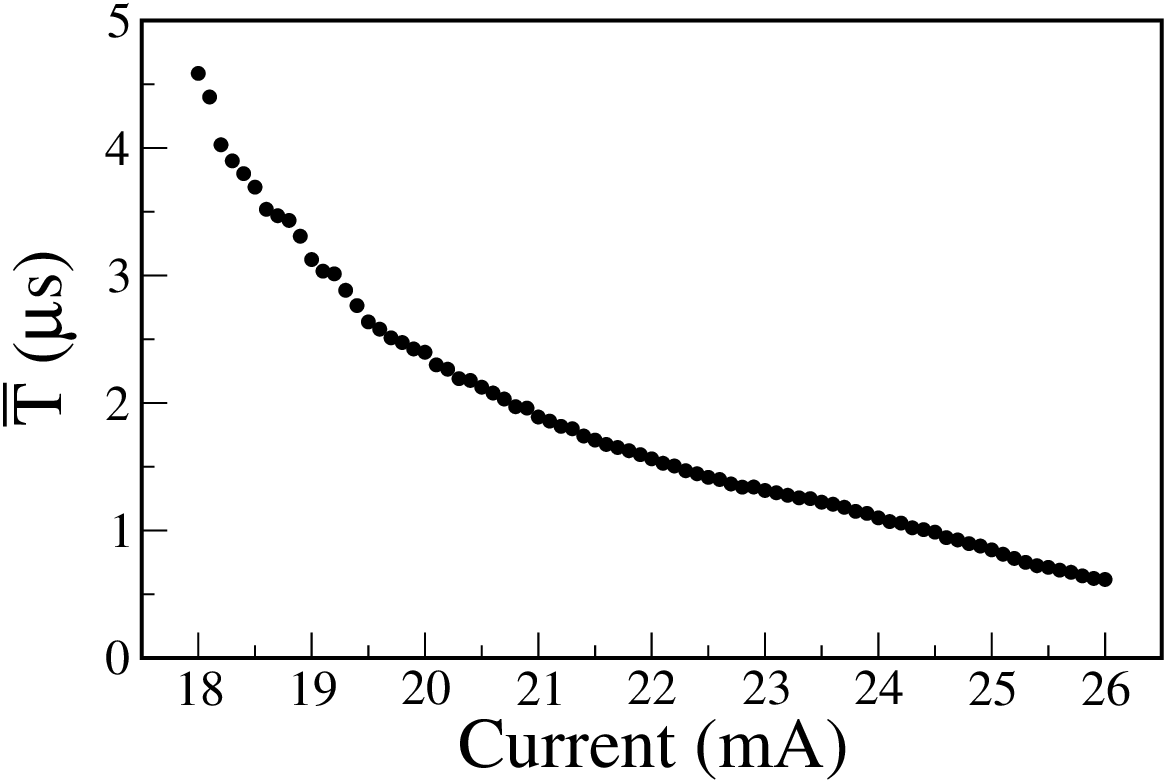}}
\caption{Experimental average of the time $T$ in Low Frequency Fluctuating power series, 
as function of the pump current.
The feedback parameter was $\xi$=12\,$\%$ and $\tau$=$30$\,ns}
\end{center}
\label{fig:temposmedio30ns}
\end{figure}

For currents just above the solitary laser threshold the time constant 
$\overline \tau_0$ shows small variation with the current and a value 
close to $3\,\tau$. Then, it decreases almost linearly with the current 
until the laser enters coherence collapse. 
The average time between drops, $T$, was also extracted from the same 
experimental data series.
The results are shown in figure \ref{fig:temposmedio30ns}. Our measured $\overline T$, 
has a value on the order of $10^2\,\tau$ for low current, and 
behaves as described in previous work \cite{liudavis}.  

\begin{figure}[!hbtp]
\begin{center}
\resizebox{7.5cm}{!}{\includegraphics{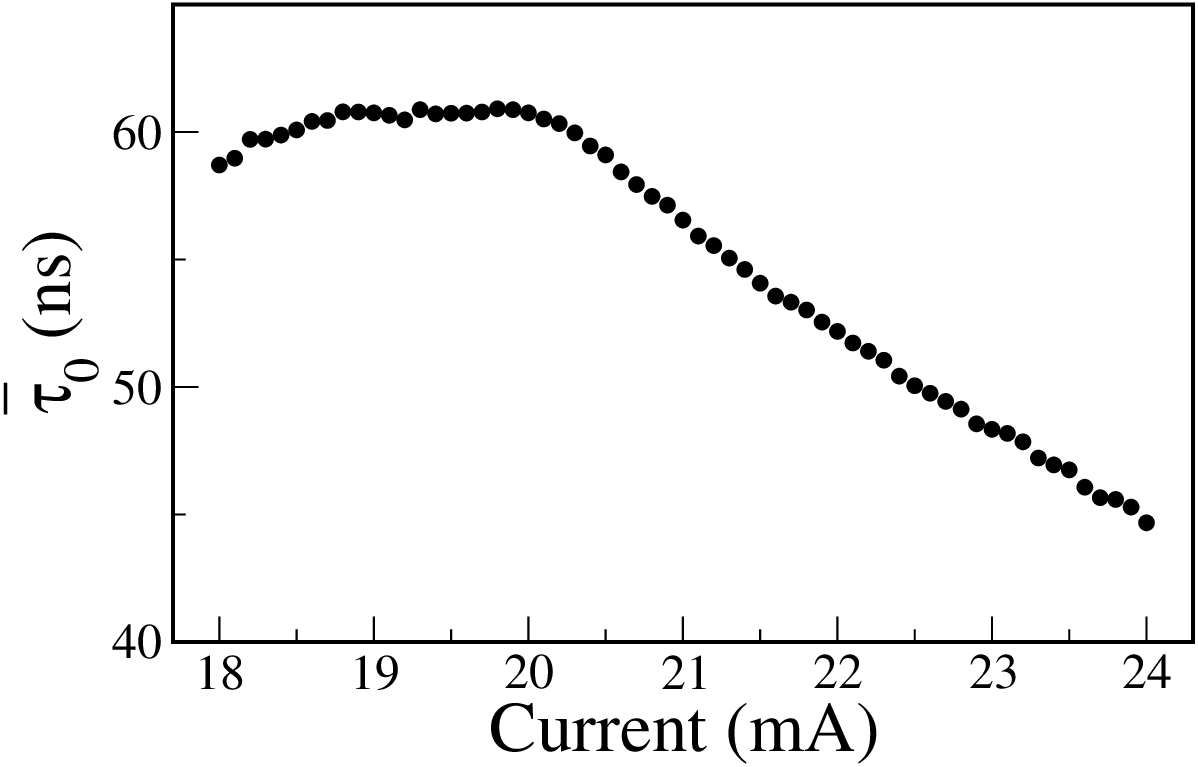}}
\caption{ Measured recovery time constant, $\tau_0$, for the diode laser with $\tau$=6ns 
delayed optical feedback and $\xi$=13.8\,$\%$.}
\end{center}
\label{fig:taumedio6ns}
\end{figure}

\begin{figure}[!hbtp]
\begin{center}
\resizebox{7.5cm}{!}{\includegraphics{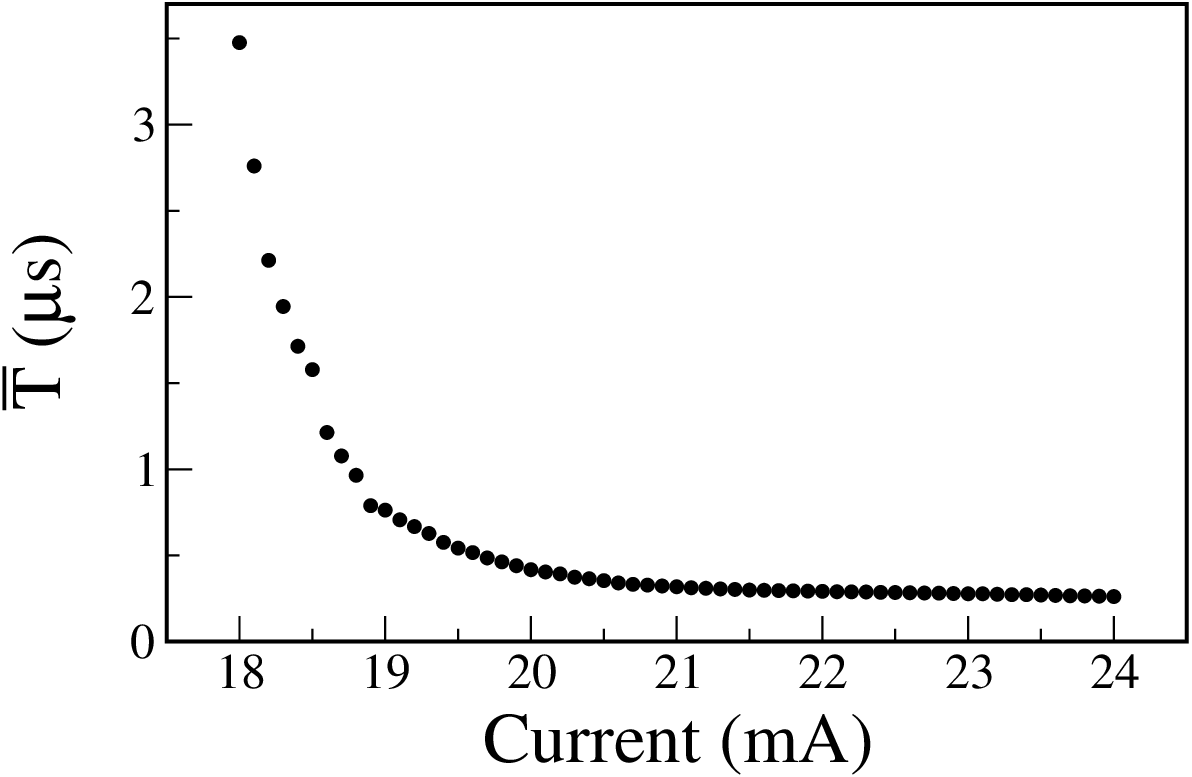}}
\caption{ Measured average time between drops, $\overline T$, 
for the diode laser with $\tau$=6ns delayed optical feedback and $\xi$=13.8\,$\%$.}
\end{center}
\label{fig:Tmedio6ns}
\end{figure}

With external cavity of 0.9 m, which corresponds to $\tau$=6 ns, the 
results obtained for $\overline \tau_0$ and $\overline T$ are shown in 
figures \ref{fig:taumedio6ns} and \ref{fig:Tmedio6ns}. 

The two quantities, $\overline T$ and $\overline \tau_{0}$, are clearly 
distinct in their dependences on the pump current. Not only their order 
of magnitude, but mainly their rate of change with the current,  given by 
the concavity of their dependences, are different. 
The most important difference appears for low currents. 
Near the solitary laser threshold current the average 
time between drops is known to have a sharp variation with the value of the 
pump current \cite{Hohlroy}.

\begin{figure}[!hbtp]
\begin{center}
\resizebox{7.5cm}{!}{\includegraphics{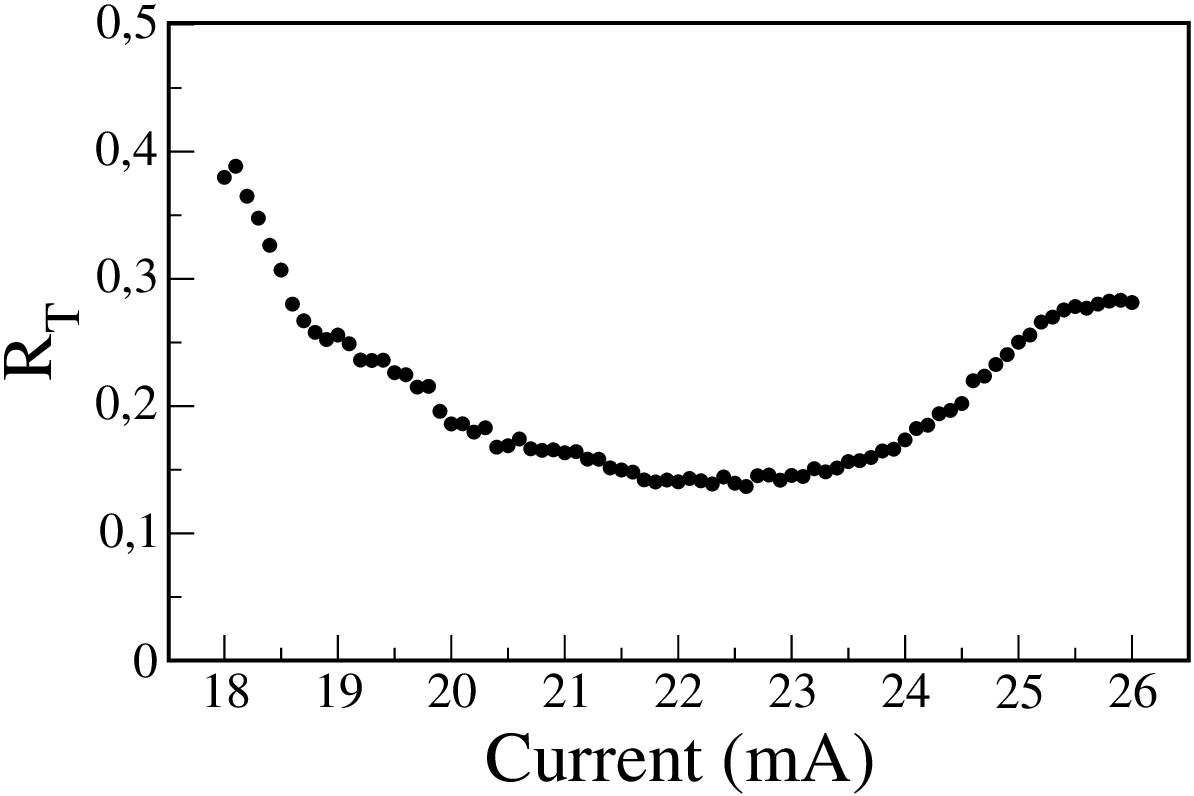}}
\caption{Experimental normalized variance of the average time $\overline T$ 
in Low Frequency Fluctuating power series,
 as function of the pump current.
The feedback parameter was $\xi$=12\,$\%$ and $\tau$=$30\,$ns}
\end{center}
\label{fig:30ld-var}
\end{figure}

\begin{figure}[!hbtp]
\begin{center}
\resizebox{7.5cm}{!}{\includegraphics{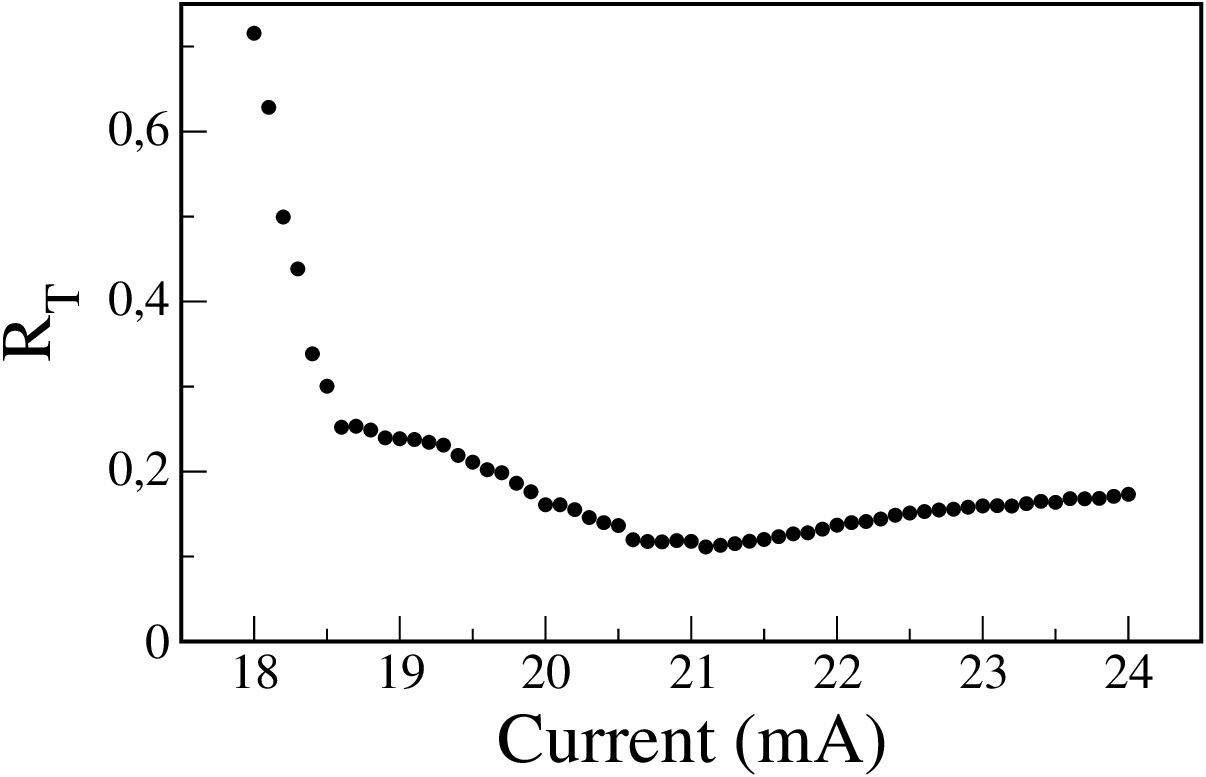}}
\caption{ Experimental relative variance of the time between drops, $T$, 
 for the diode laser with $\tau$=6ns delayed optical feedback.}
\end{center}
\label{fig:6ns-var}
\end{figure}

Further differences appear between the physical behavior of $\tau_0$ and $T$ when 
their variances are obtained as functions of the current. 
The results in figure  \ref{fig:30ld-var} and \ref{fig:6ns-var}
show the normalized variance, $R_{T}$=$\sigma_{T}/\overline T$, for 
the average time between drops. 
It has been previously studied \cite{avila2004} and
 present a minimum  for intermediate values 
 of the pump current. This can be interpreted as a manifestation of 
 deterministic coherence resonance \cite{avila2004}. 

\begin{figure}[!hbtp]
\begin{center}
\resizebox{7.5cm}{!}{\includegraphics{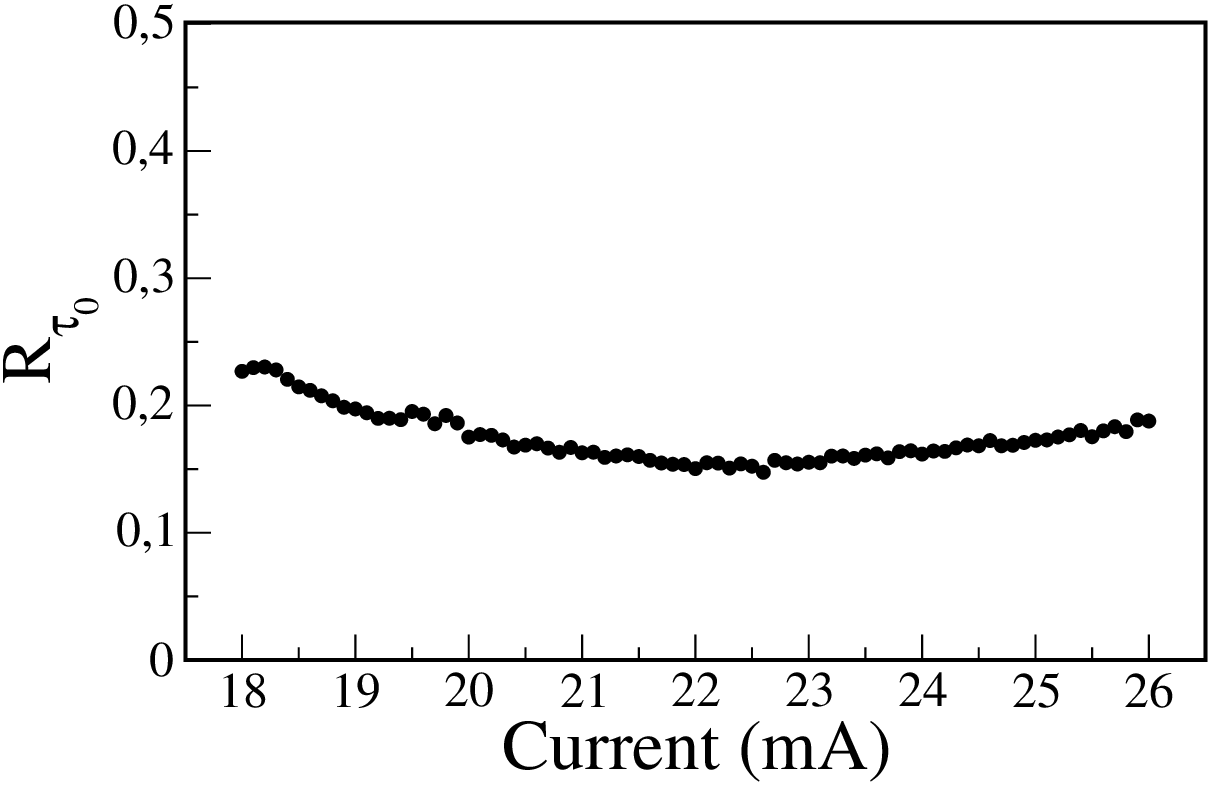}}
\caption{Experimental normalized variance of the time constant $\tau_0$ in 
Low Frequency Fluctuating power series, as function of the pump current.
The feedback parameter was $\xi$=12\,$\%$ and $\tau$=$30\,$ns.}
\end{center}
\label{fig:tau0var30ns}
\end{figure}

\begin{figure}[!hbtp]
\begin{center}
\resizebox{7.5cm}{!}{\includegraphics{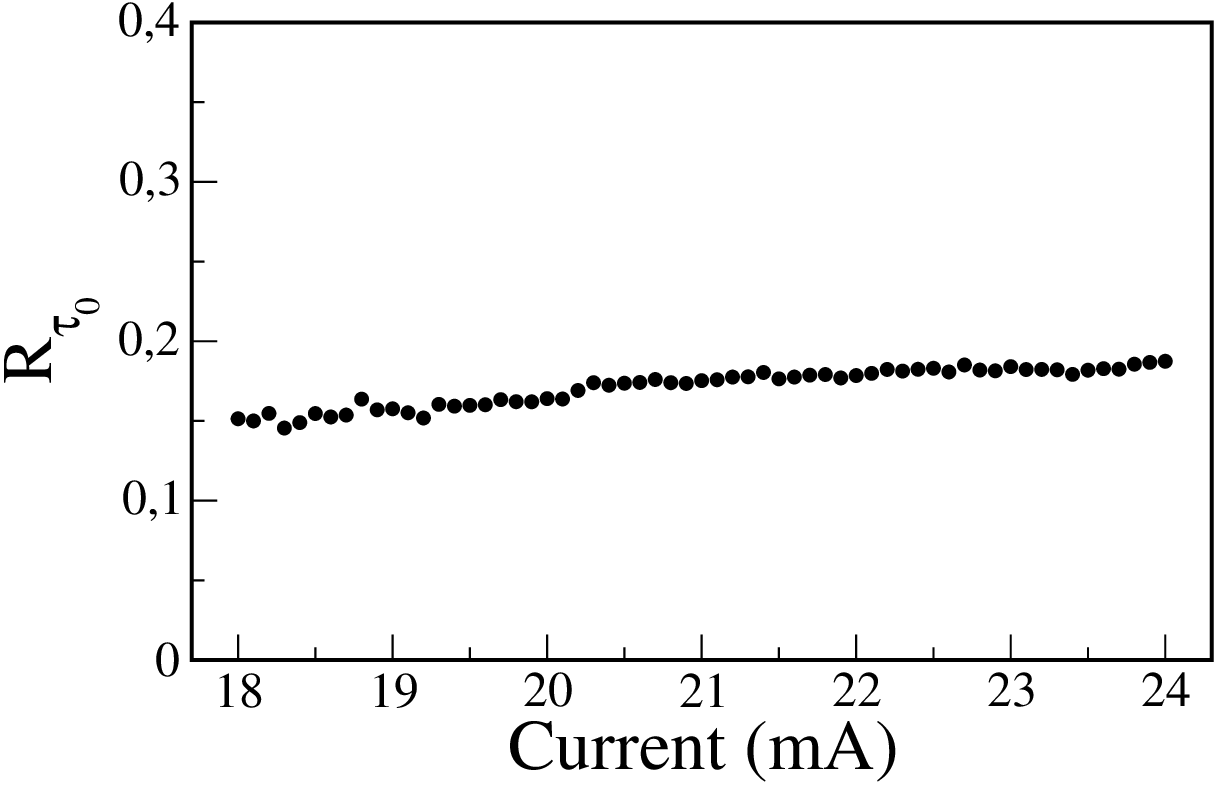}}
\caption{ Experimental  relative variance of the recovery time constant $\tau_0$ 
 for the diode laser with $\tau$=6ns delayed optical feedback.}
\end{center}
\label{fig:tau0var6ns}
\end{figure}

 The value  of ${\overline T}$  and the variance of $T$ are strongly dependent 
 on the noise in the system \cite{Hohlroy,eguia1998}. 
 For low pump currents the main noise contribution is attributed to external or 
 spontaneous emission origin \cite{Hohlroy,wing-shunroy03}, while  
  deterministic noise, resulting
  from the fast chaotic dynamics, dominates at higher currents \cite{sano,avila2004}.
  These regimes are seen in figures \ref{fig:Tmedio6ns}  and \ref{fig:6ns-var}.   
 The corresponding normalized variances $R_{\tau_{0}}=\sigma_{\tau_{0}}/\overline \tau_{0}$ 
 for the recovery time constant are shown in figure 
 \ref{fig:tau0var30ns} and 
 \ref{fig:tau0var6ns}. 
In contrast with the relative variance of $T$, the
 relative variance of the time 
constant $\tau_0$ is nearly constant. From this we infer that noise does not appear to affect
$\tau_0$. This is consistent with the definition of a quantity that is characteristic of the slow deterministic dynamics of the system.  

%Such result is taken from the experimental data 
%and is independent of 
%the nature of the noise: , with external or spontaneous emission origin, or deterministic, 
%and , the noise does not act on the recovery time constant as it does 
%on the duration of the interdrop time interval. 

Measurements done for external cavity lengths with $\tau$=6 ns, 9 ns, 15 ns, 30 ns and 60 ns  
give approximately the same behavior 
%above described properties 
for the relation between the recovery time constant and the time between drops. 
The value of $R_{\tau_0}$ remains nearly constant,
below 30\%, through the whole range of cavity lengths and pump currents investigated.  
A small concavity is observed only for relative variance of $\tau_0$ with pump current 
in the data for cavity delay time of 30 ns (figure \ref{fig:tau0var30ns}).  
For pump current near solitary threshold we have measured 
$\tau_0$ $\sim$  3$\tau$ to $\sim$  10$\tau$ and $\overline T$ $\sim$ $10^2 \tau$. 
%The nearly constant value of the variance of $\tau_0$ makes us refer to $\tau_0$ instead of the more appropriate average, $\overline \tau_0$, as we do for $T$.

\section{Conclusions}

To summarize, a new experimental quantity on the dynamics of a diode laser with optical feedback
was introduced and studied: a time constant 
 for the exponential recovery of the power drops in the Low Frequency Fluctuations.
It is defined for the LFF signal averaged on a nanosecond time scale. 
The ultrafast pulsations present 
in the LFF is filtered out  and this recovery definition  is related to the slow dynamics
(time scale longer than nanoseconds)  of the system.  
The recovery time has specific dependence on the laser parameters, as shown here for the 
measurements as function of the pump current. 
Its variance is almost constant throughout the range of currents from threshold 
up to the onset of Coherence Collapse \cite{tartwijklenstra}. 
For comparison, the well studied \cite{Hohlroy,liudavis,wing-shunroy03,sukow97,morkoc99} 
average time between drops, was also obtained for the same laser.
Our results for low pump current confirm that the recovery is very insensitive 
to the effects of noise, 
in contrast to the average time between drops. The recovery envelope  time 
constant, $\tau_0$, may be 
useful as a relevant 
measurable quantity to  be accounted 
for in calculations from theoretical models for the 
Low Frequency Fluctuations in chaotic diode lasers. A comparative study with 
numerical solutions of the Lang-Kobayahi equations \cite{langkobayashi} will 
be presented in a forthcoming publication \cite{avila2005}.    

\section*{Acknowledgments}
The authors acknowledge useful discussions with J. Tredicce 
and M. Giudici. Work partially supported by Brazilian
Agencies Conselho Nacional de Pesquisa e Desenvolvimento (CNPq) and
Funda\c{c}\~ao de Ci\^encia de Pernambuco (FACEPE) and by a Brazil-France,
Capes-Cofecub project 456/04.

\end{document}